# Improving mobility of silicon metal-oxide-semiconductor devices for quantum dots by high vacuum activation annealing

Ke Wang,[1] Hai-Ou Li,[1(a)] Gang Luo,[1] Xin Zhang,[1] Fang-Ming Jing,[1] Rui-Zi Hu,[1] Yuan Zhou,[1] He Liu,[1] Gui-Lei Wang,[2] Gang Cao,[1] Hong-Wen Jiang[3] and Guo-Ping Guo[1,4(b)]

[1] *CAS Key Laboratory of Quantum Information, University of Science and Technology of China, Hefei, Anhui 230026, China*

[2] *Key Laboratory of Microelectronics Devices & Integrated Technology, Institute of Microelectronics, Chinese Academy of Sciences, Beijing 100029, China*

[3] *Department of Physics and Astronomy, University of California, Los Angeles, California 90095, USA*

[4] *Origin Quantum Computing Company Limited, Hefei, Anhui 230026, China*

[a] Emails: haiouli@ustc.edu.cn; [b] Emails: gpguo@ustc.edu.cn

**Abstract** - To improve mobility of fabricated silicon metal-oxide-semiconductor (MOS) quantum devices, forming gas annealing is a common method used to mitigate the effects of disorder at the Si/SiO$_2$ interface. However, the importance of activation annealing is usually ignored. Here, we show that a high vacuum environment for implantation activation is beneficial for improving mobility compared to nitrogen atmosphere. Low-temperature transport measurements of Hall bars show that peak mobility can be improved by a factor of two, reaching 1.5 m$^2$/(V·s) using high vacuum annealing during implantation activation. Moreover, the charge stability diagram of a single quantum dot is mapped, with no visible disturbance caused by disorder, suggesting possibility of fabricating high-quality quantum dots on commercial wafers. Our results may provide valuable insights into device optimization in silicon-based quantum computing.

With rapid development of silicon quantum devices, quantum dots fabricated on silicon have become one of the most promising candidates for quantum computing [1]. In particular, employing nuclear spin-free $^{28}$Si [2], which is abundant in natural silicon, this kind of structure enables high fidelity control of spin qubits by offering long coherence times [3-7]. High quality silicon-based material is necessary for implementation of fault-tolerant quantum computing [8, 9]. As silicon metal-oxide semiconductors (MOS) are widely used in commercial electronic devices, specifically for their stable performances [10], leveraging complementary MOS (CMOS) technology to scale up spin-qubit manipulations is an alternative solution. Possible applications of commercial MOS technology in the field of spin qubits have been identified [11, 12]. Therefore, using commercial silicon wafers to fabricate high quality quantum dots is an important issue.

As a metric for assessing the quality of the semiconductor/oxide interface, high mobility indicates low disorder and material uniformity at the confining interfaces. Quantum dots for spin qubits usually operate at a low density, where the disorder and material uniformity affect the electrical property significantly. The increasing of the mobility suggests an enhancement of the quality of silicon wafers, which is beneficial for obtaining reproducible quantum dots [13]. Early studies on Si MOS show the results of quantum dot devices with relative low-mobility wafers [14,15]. For example, a gate-defined quantum dot has been achieved in intrinsic silicon with the mobility of 0.5 m$^2$/(V s) [16]. Researchers are trying to further improving electron mobility by reducing defect at the SiO$_2$/Si interface. A recent research shows a peak mobility of 1.4 m$^2$/(V s) and can be extended to 2.3 m$^2$/(V s) by forming gas annealing [17]. To determine the dominant sources of disorder in Si MOS devices, investigations on the magneto-transport of Hall bars of these structures were performed [18-20]. By analysing the mobility ($\mu$) as a function of density ($n$) and quantum lifetime ($\tau_q$), the dominant source limiting mobility was identified [21]. Similar experiments were performed on GaAs/AlGaAs heterostructure [22-24], Si MOSFET [25], and Si/SiGe heterostructure [26-28]. Based on these findings, several means to reduce disorder to improve mobility in Si MOS structures were found, including using strained silicon and annealing samples in N$_2$ environments [29,30].

In this Letter, we study silicon MOS Hall bars fabricated using commercial silicon wafers with different activation environments and growth methods. We obtain a peak mobility of 1.5 m$^2$/(V s), along with a critical electron density of 2.6×10$^{11}$ cm$^{-2}$, in a commercial silicon wafer by using high vacuum annealing during implantation activation in addition to forming gas annealing. The mobility of two-dimensional electron gas (2DEG) is increased by a factor of two by using activation annealing in high vacuum rather than in nitrogen atmosphere. We identify the dominant scattering mechanism that limits the peak mobility is short-range scattering, and the effect of phonon scattering can be ignored below 1.5 K. Furthermore, a single quantum dot is fabricated to demonstrate the possibility of producing high-quality quantum dots using wafers with the improved mobility.

The devices of Hall bars and quantum dots are fabricated on 8-inch commercial silicon wafers. The wafers consist of a 10-nm $SiO_2$ cap layer formed by thermal oxidation and a 725-μm Si substrate. All batches of samples are fabricated using the same process steps except for different annealing conditions. The standard fabrication process for sample preparation is as follows: ion implantation (phosphorus) with a dose of $2 \times 10^{15}$ cm$^{-2}$; application of an annealing process for ion activation at the temperature of 1050 ℃ in either high vacuum $(<10^{-3}$ Pa$)$ or $N_2$ atmosphere; fabrication of Ohmic contact regions using buffered oxide etching and evaporation of Ti/Au; and deposition of 30-nm-thick aluminium oxide ($Al_2O_3$) as gate insulator at the temperature of 300 ℃. After that, a layer of Al above the $Al_2O_3$ is added to define the conductive channel. A final annealing with forming gas (15% $H_2$ and 85% $N_2$) for 30min at the temperature of 400 ℃ is performed to reduce the defects induced by e-beam lithography and metal evaporation [17, 31].

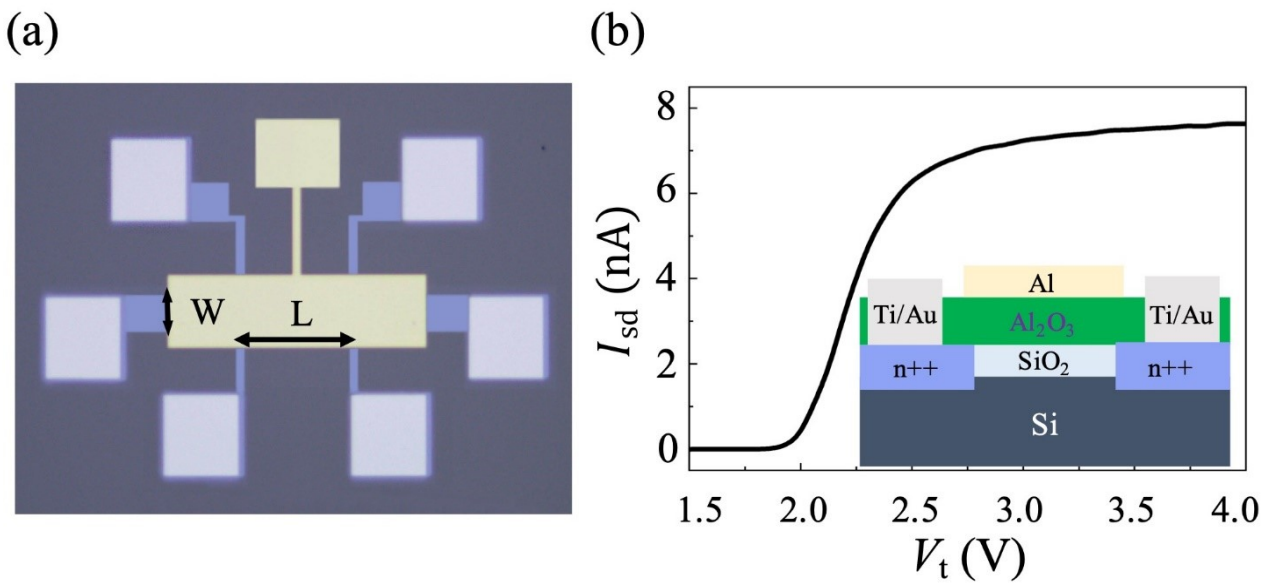

Fig. 1: (Colour online) (a) Optical micrograph of a Si MOS Hall bar showing the ion implantation region (grey areas) along with Ti/Au pads (white squares) and the top gate (faint yellow area) made of Al. (b) Standard current-voltage

curve obtained under a source-drain bias of 20 μV. Inset: Cross section of the Si MOS Hall bar device.

A channel length of L = 300 μm and width of W = 100 μm are chosen for the Hall bars on commercial silicon wafers [Fig. 1(a)]. The inset of Fig. 1(b) shows a cross section schematic of the device. The top gate (faint yellow area) is for applying a DC voltage $V_t$ to accumulate the 2DEG at the Si/SiO$_2$ interface to generate a current with a voltage $V_t > 1.89$ V. The standard 'turn on' curve of the current (Fig. 1(b)) saturates at $V_t \sim 3$ V with a total resistivity of 2.5 kΩ at 4.2 K.

Then, low-temperature transport measurements are performed at temperatures 0.24 K < T < 1.5 K in a He-3 refrigerator using an SR830 lock-in amplifier at a frequency of ~70.23 Hz to reveal the properties of our devices. In our devices, the density of the 2DEGs shows a linear dependence on the top gate voltage, with a typical curve shown in Fig. 2(a). Meanwhile, the mobility of the 2DEGs also changes as the gate voltage varies. The dependence of the mobility ($\mu$) on the density ($n$) under different top gate voltages is presented in Fig. 2(b). In the low density ($n <$ 6×10$^{11}$ cm$^{-2}$) regime, the curve roughly follows a $\mu \propto n^3$ trend, corresponding to the effect of fixed charges [32]. The mobility decreases when further increasing the density ($n > 1×10^{12}$ cm$^{-2}$) due to interface-roughness scattering at the Si/SiO$_2$ interface [33]. The mobility reaches a peak value of 1.5 m$^2$/(V·s) at a density of 9.06×10$^{11}$ cm$^{-2}$

In Fig. 2(c), at the peak mobility, Hall transport measurements are performed by mapping the longitudinal resistivity $\rho_{xx}$ and the transverse resistivity $\rho_{xy}$ under different magnetic fields. The onset of magnetic field for Shubnikov-de Hass (SdH) oscillations is 0.7 T ($B_{SdH}$). When the magnetic field is above $B_{SdH}$, the transverse resistivity $\rho_{xy}$ shows plateaus with an even filling factor $\nu =$ 4, 6, 8,… while oscillations of the longitudinal resistivity $\rho_{xx}$ are observed. The transverse resistivity $\rho_{xy}$ plateaus with odd filling-factor are invisible as a result of valley degeneracy. Spins split at a higher magnetic field of $B_s = 3.1$ T ($\nu = 10$), corresponding to a Zeeman splitting energy of $E_z = g\mu_B B_s =$ 360 μeV, where $g = 2$ is the electronic Lande factor and $\mu_B$ is the Bohr magneton.

In Fig. 2(d), we obtain $\rho_{xx}$ as a function of magnetic field ($B$) from 1 T to 2 T. After subtracting the polynomial background and $\rho_{xx}(0)$ (the longitudinal resistivity at zero magnetic field), six oscillations $\Delta\rho_{xx}(B) = \rho_{xx}(B) - \rho_{xx}(0)$ versus $1/B$ are plotted under different temperatures. The oscillations show a period of 4 in $\nu$, consistent with two-fold valley degeneracy and two-fold spin degeneracy. The oscillations can be fitted perfectly using the formula [34]

$$\Delta\rho_{xx}(B) = 4\rho_{xx}(0) \cdot \cos(\frac{E_F}{\hbar\omega_c}) \cdot D(m^*,T) \cdot \exp(-\pi/\omega_c\tau_q), \tag{1}$$

where $D(m^*,T)$ is the Dingle factor, where at low field, $D(m^*,T) = \xi/\sinh\xi$, with $\xi = 2\pi^2 k_B T / \hbar\omega_c$, $\omega_c = eB/m^*$ is the cyclotron frequency, $m^*$ is the effective mass, $k_B$ is Boltzmann's constant, $\tau_q$ is the quantum lifetime of the electrons and $\hbar$ is the reduced Planck constant. When $\xi \gg 1$, the extrapolated amplitude for $\Delta\rho$ in Eq. (1) can be used to obtain effective mass $m^* = 0.19 m_0$, which is consistent with the transverse effective mass of silicon in theory.

Figure 2(e) shows $\ln\left(\Delta\rho_{xx}(B)\cdot\sinh\xi/\xi\right)$ as a function of $1/B$, from which the quantum lifetime values $\tau_q$ can be extracted. We find $\tau_q$ is independent with temperature, similar to the transport lifetime $\tau_t = \mu \cdot m^*/e$, as shown in Fig. 2(f). We note that this result differs from previous works on Si/SiGe heterostructure [28] and GaAs/AlGaAs heterostructures [24], where phonon scattering was nonnegligible. In addition, the value of $\tau_q$=0.8 ps implies a Landau-level broadening of $\Gamma = \hbar/2\tau_q = 410$ µeV, which is consistent with the reported value of 480 µeV [13], giving the upper bound of valley splitting energy in our devices. For comparison, valley splitting energies range from 0.7 to 1.5 meV in previous Si MOSFETs [35-38].

It is worth noting that the Dingle ratio $= \tau_q / \tau_t$ almost equals unity, indicating that the dominant scattering mechanism restricting the peak mobility of this sample [Fig. 2(f)] is short-range scattering rather than long-range scattering. This result is in accordance with previous work in which the fixed charges and defects at the SiO$_2$/Al$_2$O$_3$ or SiO$_2$/Si interfaces dominate the scattering [39]. The insensitivity of both $\tau_q$ and $\tau_t$ to temperature implies that phonon effects are not dominant here even if we cannot rule out other effects that limit the mobility of the wafers [40].

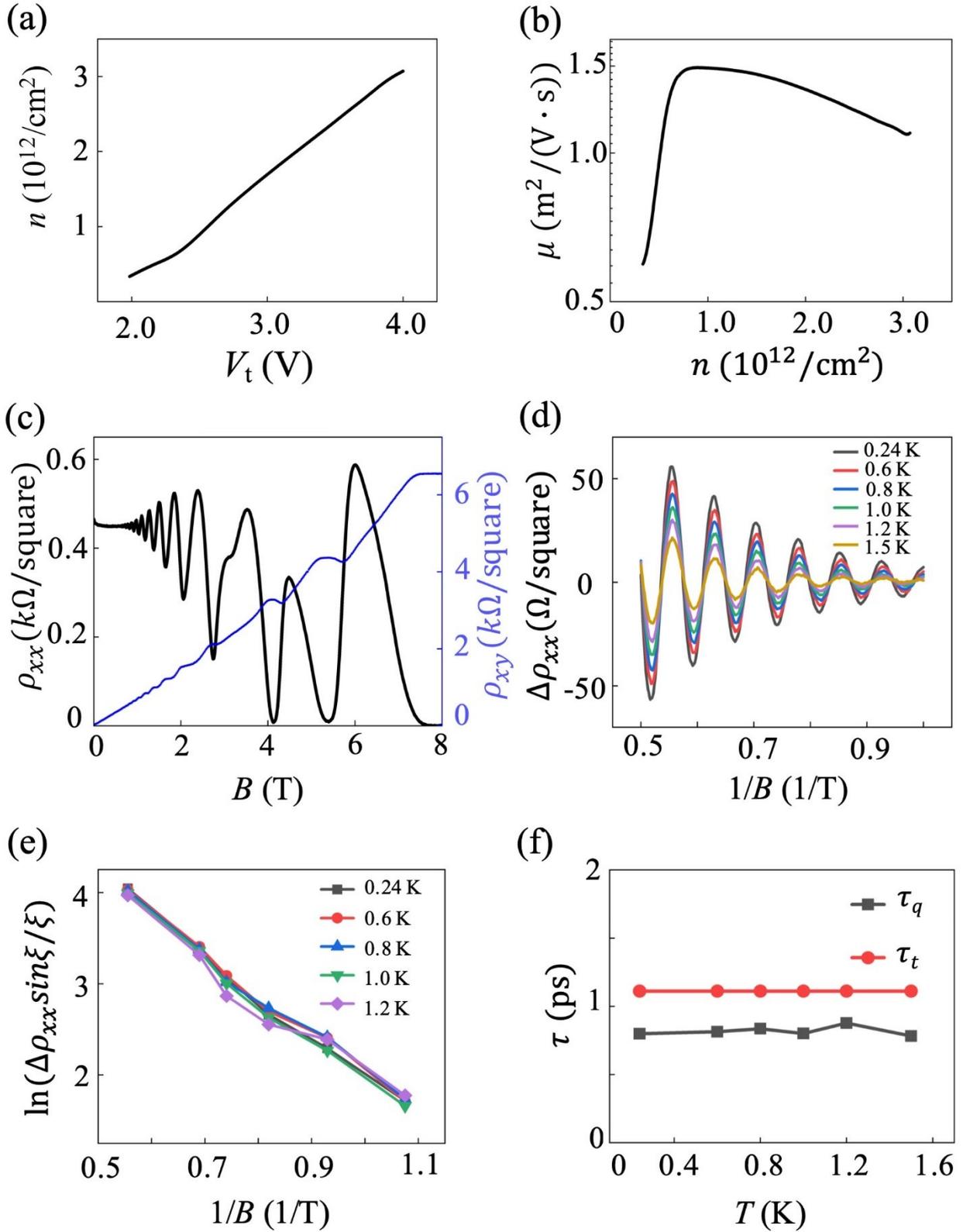

Fig. 2: (Colour online) (a) 2DEG Hall density $n$'s dependency on top gate voltage $V_t$. (b) Mobility versus density of device of batch 3 (high vacuum). (c) Longitudinal resistivity ($\rho_{xx}$, black) and transverse resistivity ($\rho_{xy}$, blue) at $n$ = 9.06×10$^{11}$ cm$^{-2}$ as a function of magnetic field. (d) Fourier spectrum of SdH oscillations from 1 T to 2 T in the temperature range of $T = 0.24 - 1.5$ K after subtracting the polynomial background and $\rho_{xx}$ (0). (e) Magnetic field dependence of the logarithm of $\Delta\rho_{xx}$ under different temperatures. (f) Quantum lifetime $\tau_q$ and transport lifetime $\tau_t$ are independent of temperature.

In Fig. 3(a), by varying the gate voltage $(V_t)$, $\rho_{xx}$ ($B$=0 T) as a function of density ($n$) is obtained. After changing the temperature, these curves feature a stable crossing point at a critical electron density ($n_c = 2.6 \times 10^{11}$ cm$^{-2}$), which is a signature of metal-insulator transition. The value of $n_c$ in our MOS Hall bars is consistent with the previously reported value [41-44]. In Fig. 3(b), the peak mobility and density for different parameter settings are listed. Data of hollow symbols are from devices using high vacuum ($<10^{-3}$ Pa) activation annealing, while those of solid symbols come from devices activated in nitrogen atmosphere. Moreover, different wafers, which can be used for quantum dots fabrication, are introduced to identify if there is any difference between intrinsic substrates (Floating zone-FZ grown wafers) and doped substrates (Czochralski-Cz grown wafers), even both of them can be used for quantum dots fabrication [45-47]. Batches 1 and 2 are fabricated using Cz wafers, while batch 3 is fabricated with FZ wafers. Cz wafers are slightly boron (B+)-doped with a resistivity of 20–30 Ω·cm, whereas FZ wafers are intrinsic and undoped with a resistivity of approximately 10 kΩ·cm. Comparing ones with same wafer but different annealing conditions, a higher peak mobility is clearly observed for the ones under high vacuum activation annealing instead of nitrogen atmosphere annealing. A peak mobility up to 1.5 m$^2$/(V·s) is obtained for batch 3. Overall, the peak mobility of the 2DEGs fabricated using FZ wafers is higher than that of the 2DEGs fabricated using Cz wafers here. We attribute this trend to fewer donors in FZ wafers than in Cz wafers. Considering dopant atoms, the background doping values of Cz wafers and FZ wafers are approximately 10$^{14}$ cm$^{-3}$ and 10$^{12}$ cm$^{-3}$, respectively. For FZ wafers, the possible effects from dopant atoms are relatively small. Additionally, for quantum dots in Si MOS, three kinds of charge transitions are usually considered, namely, E' centres, P$_b$ centres, and unintentional quantum dots [39]. For E' centres, the effect is related to the fixed charges, typically arising from the interface between Al$_2$O$_3$ and SiO$_2$. This kind of defect is usually induced by radiation during e-beam evaporation and is usually located further from the quantum well. Such defects could be worsened when implementing e-beam lithography for smaller structures [31]. Another kind of defect, the P$_b$ centre, is formed by a single dangling bond on silicon [48]. Previous experiments have identified that the most common defects of the P$_b$ centre form at the interface between Si and SiO$_2$ [49]. The final annealing with H$_2$ is performed to passivate the P$_b$ centres.

Unintentional quantum dots are formed by strains between different materials [50], which is not discussed here. All of the effects would behave as disorders and they usually occur simultaneously. Mitigation of any of these effects will lead to an increase of electron mobility.

In Fig. 3(c), we focus on electron mobility's dependency on density for both high vacuum annealing activated and nitrogen atmosphere annealing activated devices. Taking batch 1 as an example, the device of high vacuum activation annealing has a higher mobility. The mobility at the same density is increased by a factor of two by replacing nitrogen atmosphere annealing with high vacuum activation annealing. The results in Fig. 3 (d) show that our improvement by high vacuum annealing is universal for both intrinsic substrates and boron doped substrates. Additionally, samples of batch 3 have higher mobilities than those of batches 1 and 2 for both two methods, which may possibly be attributed to strain releasing and diffusing of impurity atoms induced by high vacuum activation annealing. Overall, to achieve a higher peak mobility, high vacuum activation annealing is a promising solution and FZ substrate is premium.

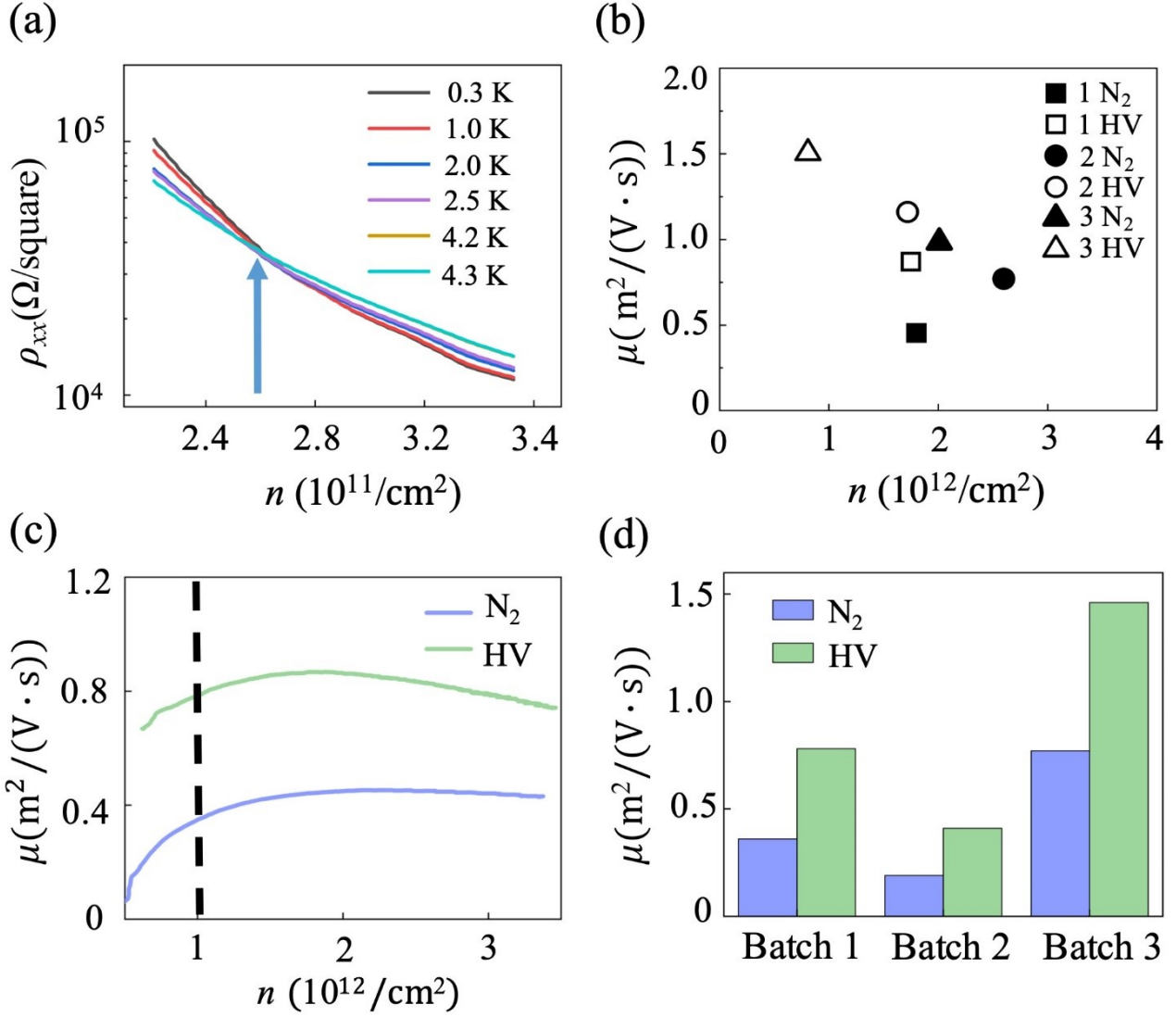

Fig. 3: (Colour online) (a) Resistivity $\rho_{xx}$ as a function of density $n$ in the temperature range of $T = 0.3 - 4.3$ K. (b) Comparison of the peak mobility and density from different batches of devices. Batch 1: Cz silicon wafers with a layer of home-grown silicon oxide. Batch 2: Cz commercial silicon/silicon oxide wafers. Batch 3: FZ commercial silicon/silicon oxide wafers. Data of hollow symbols are from devices whose activation is implemented in a high vacuum (HV) environment, solid symbols correspond to devices activated in nitrogen atmosphere. (c) Electron mobility as a function of density obtained from a device belongs to batch 1. (d) Comparison of the mobility differences at a density of $1 \times 10^{12}$ cm$^{-2}$ between high vacuum annealing activated and nitrogen atmosphere annealing activated devices.

We demonstrate a single quantum dot using Cz wafers (batch 2) with a improved mobility of 1.2 m$^2$/(V·s) by high vacuum activation annealing and the result shows the commercial wafers after our processing can be used for fabricating quantum dot devices. The quantum dot design was based on a double top-gated architecture, similar to previous work conducted on GaAs and Si [51,52]. This single quantum dot and quantum point contact structure is confined by six gates between the $Al_2O_3$ layer and $SiO_2$ layer. The structure of the planar gates is displayed in Fig. 4(a). By detecting the charge sensing signal of the single dot, the standard stability diagram of the single dot can be obtained. In Fig. 4 (b), dozens of tunnelling lines are mapped by changing the gate voltages at G3 and G5. These parallel lines show a single dot behaviour. Moreover, the interval between two proximal lines gradually increases when lowering the gate voltages, corresponding to the change in the electron number in the dot. Clearly, these lines are not cut off by other lines arising from disorder. No impurities are involved in this regime. More than 15 consecutive Coulomb oscillations can be observed without visible parasitic dot effect. The collection of process described in the paper to improve the device mobility, in a long run, should help to improve the quality of Si MOS quantum dots.

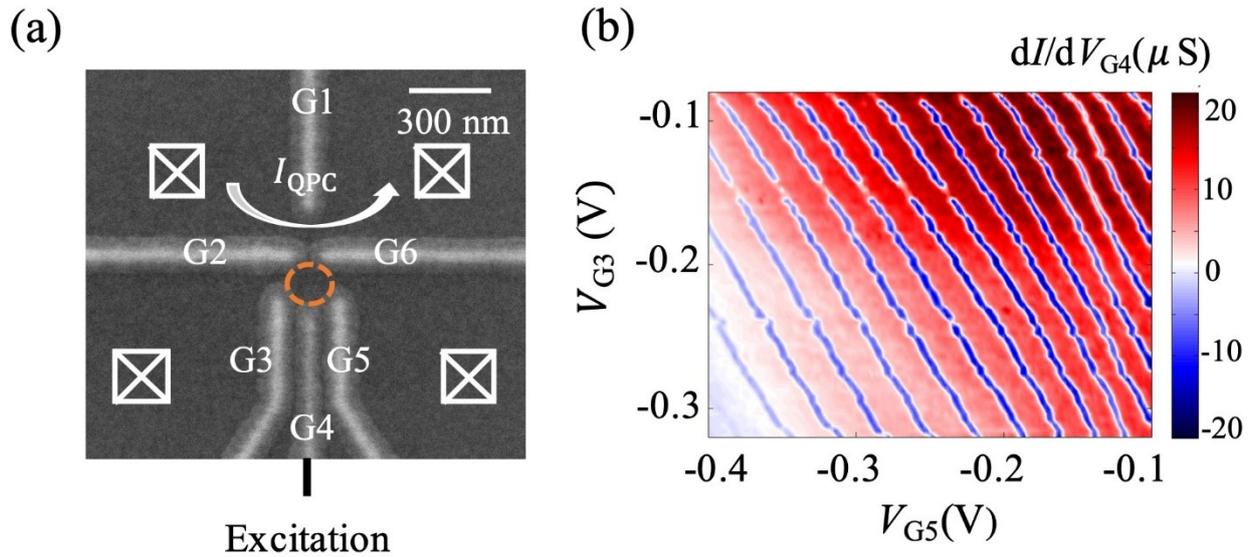

Fig. 4: (Colour online) (a) Scanning electron microscopy image of a single quantum dot and quantum point contact (QPC) structure with buried oxide on a Si MOS heterostructure. G1–G6 are the depleted gates used to define the potential of the quantum well. Excitation is a sinusoidal waveform coming from a lock-in for charge sensing. (b) Charge stability diagram of the single quantum dot monitored via an adjacent quantum point contact. Each line corresponds to a tunnelling event of electrons.

In conclusion, we perform the quantum transport characterizations on silicon MOS Hall bars using commercial silicon wafers. We report that high vacuum activation annealing helps to increase the mobility of 2DEGs by a factor of two, leading to a peak mobility of $1.5\ m^2/(V \cdot s)$. We find that high vacuum activation annealing and selection of FZ wafers are helpful to approach the higher mobility.

And the behavior of our single dot device is a good example of improving the mobility for high quality quantum dot formation. For further improvement in quantum devices, more efforts are still needed to handle residual impurities and defects at the interface between Si and $SiO_2$. Our results may provide valuable insights into wafer optimization in silicon-based quantum computing.

<div align="center">***</div>

This work was supported by the National Key Research and Development Program of China (Grant No. 2016YFA0301700), the National Natural Science Foundation of China (Grant Nos. 61674132, 11674300 and 11625419), the Strategic Priority Research Program of the CAS (Grant No. XDB24030601), and the Anhui Initiative in Quantum Information Technologies (Grant No. AHY080000), and this work was partially carried out at the USTC Center for Micro and Nanoscale Research and Fabrication.

## References

[1] Zwanenburg F. A., Dzurak A. S., Morello A., Simmons M. Y., Hollenberg L. C. L., Klimeck G., Rogge S., Coppersmith S. N. and Eriksson M. A., *Rev. Mod. Phys.*, **85** (2013) 961.

[2] Itoh K. M. and Watanabe H., *Mrs Commun.*, **4** (2014) 143.

[3] Veldhorst M., Yang C. H., Hwang J. C., Huang W., Dehollain J. P., Muhonen J. T., Simmons S., Laucht A., Hudson F. E., Itoh K. M., Morello A. and Dzurak A. S., *Nature*, **526** (2015) 410.

[4] Zajac D. M., Sigillito A. J., Russ M., Borjans F., Taylor J. M., Burkard G. and Petta J. R., *Science*, **359** (2018) 439.

[5] Watson T. F., Philips S. G. J., Kawakami E., Ward D. R., Scarlino P., Veldhorst M., Savage D. E., Lagally M. G., Friesen M., Coppersmith S. N., Eriksson M. A. and Vandersypen L. M. K., *Nature*, **555** (2018) 663.

[6] Yoneda J., Takeda K., Otsuka T., Nakajima T., Delbecq M. R., Allison G., Honda T., Kodera T., Oda S., Hoshi Y., Usami N., Itoh K. M. and Tarucha S., *Nat Nanotechnol.*, **13** (2018) 102.

[7] Zhang X., Li H. O., Cao G., Xiao M., Guo G. C. and Guo. G. P., *Natl Sci Rev.*, **6** (2019) 32-54.

[8] Fowler A. G., Mariantoni M., Martinis J. M. and Cleland A. N., *Phys. Rev. A*, **86** (2012) 032324.

[9] Hill C. D., Peretz E., Hile S. J., House M. G., Fuechsle M., Rogge S., Simmons M. Y. and Hollenberg L. C. L., *Science Advances*, **1** (2015) e1500707.

[10] Vandersypen L. M. K., Bluhm H., Clarke J. S., Dzurak A. S., Ishihara R., Morello A., Reilly D. J., Schreiber L. R. and Veldhorst M., *Npj Quantum Inform.*, **3** (2017) 34.

[11] Maurand R., Jehl X., Kotekar-Patil D., Corna A., Bohuslavskyi H., Lavieville R., Hutin L., Barraud S., Vinet M., Sanquer M. and De Franceschi S., *Nat Commun.*, **7** (2016) 13575.

[12] Barraud S., Coquand R., Casse M., Koyama M., Hartmann J. M., Maffini-Alvaro V., Comboroure C., Vizioz C., Aussenac F., Faynot O. and Poiroux T., *Ieee Electr Device L*, **33** (2012) 1526.

[13] Sabbagh D., Thomas N., Torres J., Pillarisetty R., Amin P., George H. C., Singh K., Budrevich A., Robinson M.,

Merrill D., Ross L., Roberts J., Lampert L., Massa L., Amitonov S. V., Boter J. M., Droulers G., Eenink H. G. J., van Hezel M., Donelson D., Veldhorst M., Vandersypen L. M. K., Clarke J. S., and Scappucci G., *Phys. Rev. Applied*, **12** (2019) 014013.

[14] Jupina M. A. and Lenahan P. M., *IEEE Trans. Nucl. Sci.*, **37** (1990) 1650.

[15] Eng, K., McFarland, R. N., and Kane, B. E. (2007). *Phys. Rev. Lett*, **99** (2007) 016801.

[16] Angus, S. J., Ferguson, A. J., Dzurak, A. S., and Clark, R. G. *Nano Lett*. **7** (2007) 2051-2055.

[17] Kim J. S., Tyryshkin A. M. and Lyon S. A., *Appl. Phys. Lett.*, **110**, (2017) 123505.

[18] Bianchi R. A., Bouche G. and Roux-dit-Buisson O., *International Electron Devices 2002 Meeting, Technical Digest* 117.

[19] Wang M. C., Yang H. C. and Liao W. S., *Adv Mater Res-Switz*, **287** (2011) 2974.

[20] Saks N. S., Klein R. B. and Griscom D. L., *IEEE Trans. Nucl. Sci.*, **35** (1988) 1234.

[21] Das Sarma S. and Stern F. *Phys. Rev. B*, **32** (1985) 8442.

[22] Jiang C., Tsui D. C. and Weimann G., *Appl. Phys. Lett.*, **53** (1988) 1533.

[23] Coleridge P. T., *Phys. Rev. B*, **44** (1991) 3793.

[24] Tarquini V., Knighton T., Wu Z., Huang J., Pfeiffer L. and West K., *Appl. Phys. Lett.*, **104** (2014) 092102.

[25] Chain K., Huang J. H., Duster J., Ko P. K. and Hu C. M., *Semicond Sci Tech.*, **12** (1997) 355.

[26] Li J. Y., Huang C. T., Rokhinson L. P. and Sturm J. C., *Appl. Phys. Lett.*, **103** (2013) 162105.

[27] Ismail K., Arafa M., Saenger K. L., Chu J. O. and Meyerson B. S., *Appl. Phys. Lett.*, **66** (1995) 1077.

[28] Mi X., Hazard T. M., Payette C., Wang K., Zajac D. M., Cady J. V. and Petta J. R., *Phys. Rev. B*, **92** (2015) 035304.

[29] Rim K. K., Hoyt J. L. and Gibbons J. F., *IEEE Trans. Electron Devices*, **47** (2000) 1406.

[30] Hoyt J. L., Nayfeh H. M., Eguchi S., Aberg I., Xia G., Drake T., Fitzgerald E. A. and Antoniadis D. A., *International Electron Devices 2002 Meeting, Technical Digest* 23.

[31] Kim J. S., Hazard T. M., Houck A. A. and Lyon S. A., *Appl. Phys. Lett.*, **114**, (2019) 043501.

[32] Ando T., Fowler A. B. and Stern F., *Rev. Mod. Phys.* **54** (1982) 437.

[33] Gold A., *Phys. Rev. Lett.*, **54** (1985) 1079.

[34] Ando T. and Uemura Y., *J. Phys. Soc. Jpn.* **36** (1974) 959.

[35] Pudalov V. M., Semenchinskii S. G. and Edelman V. S., *Jetp Lett+*. **41** (1985) 325.

[36] Takashina K., Ono Y., Fujiwara A., Takahashi Y. and Hirayama Y., *Phys. Rev. Lett.*, **96** (2006) 236801.

[37] Kohler H. and Roos M., *Physica Status Solidi B-Basic Research*, **91** (1979) 233.

[38] Yang, C. H. and Rossi, A. and Ruskov, R. and Lai, N. S. and Mohiyaddin, F. A. and Lee, S. and Tahan, C. and Klimeck, G. and Morello, A. and Dzurak, A. S. *Nat Commun*., **4** (2013) 2069.

[39] Spruijtenburg P. C., Amitonov S. V., Mueller F., van der Wiel W. G. and Zwanenburg F. A., *Sci Rep.*, **6** (2016) 38127.

[40] Das Sarma S., Hwang E. H., Kechedzhi K. and Tracy L. A., *Phys. Rev. B*, **90** (2014) 125410.

[41] Popovic D., Fowler A. B. and Washburn S., *Phys. Rev. Let.*, **79** (1997) 1543.

[42] Kravchenko S. V., Kravchenko G. V., Furneaux J. E., Pudalov V. M. and Diorio M., *Phys. Rev. B*, **50** (1994) 8039.

[43] Morello A., Pla J. J., Zwanenburg F. A., Chan K. W., Tan K. Y., Huebl H., Mottonen M., Nugroho C. D., Yang C. Y., van Donkelaar J. A., Alves A. D. C., Jamieson D. N., Escott C. C., Hollenberg L. C. L., Clark R. G. and Dzurak A. S., *Nature*, **467** (2010) 687.

[44] Zhu W. J., Han J. P. and Ma T. P., *Electron Devices*, **51** (2004) 98.

[45] Petit, L. and Boter, J. M. and Eenink, H. G. J. and Droulers, G. and Tagliaferri, M. L. V. and Li, R. and Franke, D. P. and Singh, K. J. and Clarke, J. S. and Schouten, R. N. and Dobrovitski, V. V. and Vandersypen, L. M. K. and Veldhorst, M., *Phys. Rev. Lett.* **121** (2018) 076801.

[46] Rochette, S. and Rudolph, M. and Roy, A.-M. and Curry, M. J. and Eyck, G. A. Ten and Manginell, R. P. and Wendt, J. R. and Pluym, T. and Carr, S. M. and Ward, D. R. and et al. *Appl. Phys. Lett.*, **114**, (2019) 083101.

[47] Nordberg, E. P. and Eyck, G. A. Ten and Stalford, H. L. and Muller, R. P. and Young, R. W. and Eng, K. and Tracy, L. A. and Childs, K. D. and Wendt, J. R. and Grubbs, R. K. and Stevens, J. and Lilly, M. P. and Eriksson, M. A. and Carroll, M. S., *Phys. Rev. B* **80** (2009) 115331.

[48] Thoan N. H., Keunen K., Afanas'ev V. V. and Stesmans A. *J. Appl. Phys.*, **109** (2011) 013710.

[49] Eng K., McFarland R. N. and Kane B. E. *Appl. Phys. Lett.* **87** (2005) 052206.

[50] Spruijtenburg P. C., Amitonov S. V., van der Wiel W. G., et al. *Nanotechnology*, **29** (2018) 143001.

[51] Li H. O., Cao G., Xiao M., You J., Wei D., Tu T., Guo G. C., Jiang H. W. and Guo G. P., *J. Appl. Phys.*, **116** (2014) 174504.

[52] Xiao M., House M. G. and Jiang H. W., *Phys. Rev. Lett.* **104** (2010) 096801.